\documentclass[nofootinbib,aps,a4paper,letterpaper,superscriptaddress,
twocolumn,times,eqsecnum,longbibliography]{revtex4}
\usepackage{amsmath}
\usepackage{amsfonts}
\usepackage{booktabs}
\usepackage{multirow}
\usepackage{siunitx} 
\usepackage{adjustbox}
\usepackage{graphicx}
\usepackage{color}
\usepackage{braket}
\usepackage{dcolumn}
\usepackage{bm,url}
\usepackage[linktocpage]{hyperref}
\usepackage{subfigure}
\usepackage{amsfonts}
\usepackage{orcidlink}
\usepackage[usenames,dvipsnames,svgnames]{xcolor}
\usepackage{hyperref}   
\definecolor{oxfordblue}{rgb}{0.0, 0.13, 0.28}
\definecolor{burgundy}{rgb}{0.5, 0.0, 0.13}
\definecolor{darkolivegreen}{rgb}{0.33, 0.42, 0.18}
\definecolor{darkblue}{rgb}{0,0,0.5}
\definecolor{richcarmine}{rgb}{0.84, 0.0, 0.25}
\definecolor{bluer}{rgb}{0.00,0.50,0.75}{}
\hypersetup{colorlinks=true, citecolor=red, linkcolor=blue,
 urlcolor = magenta, filecolor=magenta}

\begin{document}

\title{Area-product universality in multi-horizon black hole entropy}

\author{Francisco Tello-Ortiz
\orcidlink{0000-0002-7104-5746}
}
\email{francisco.tello@ufrontera.cl}
\affiliation{Departamento de Ciencias F\'isicas, Universidad de La Frontera, Casilla 54-D, 4811186 Temuco, Chile.}
 
\author{Y. G\'omez-Leyton\orcidlink{0000-0003-3385-4785}}
\email{ygomez@ucn.cl}
\affiliation{Departamento de F\'isica, Universidad Cat\'olica del Norte, Av. Angamos 0610, Antofagasta, Chile.}

\author{Jean B\'aez Cuevas\orcidlink{0000-0002-3308-3362}}
\email{jean.baez@pucv.cl}
\affiliation{
Instituto de F\'isica, Pontificia Universidad Cat\'olica de Valpara\'iso, Casilla 4950, Valpara\'iso, Chile.}

 \author{Emmanuel N. Saridakis\orcidlink{0000-0003-1500-0874}}
 \email{msaridak@noa.gr}
 \affiliation{The National Observatory of Athens, Lofos Nymfon 11852, Greece}
 \affiliation{Departamento de Matem\'{a}ticas, Universidad Cat\'{o}lica del 
  Norte, Avda. Angamos 0610, Casilla 1280, Antofagasta, Chile}
 \affiliation{CAS Key Laboratory for Research in Galaxies and Cosmology, 
School   of Astronomy and Space Science,
  University of Science and Technology of China, Hefei 230026, China}

\begin{abstract}
Black-hole entropy and multi-horizon area-product relations represent two 
apparently distinct manifestations of universality, arising respectively from 
microscopic statistics and classical geometry. We show that these structures 
are 
unexpectedly connected. Extending a minimum-assumptions statistical 
discretization to multiple Killing horizons, with distinct horizon sectors 
treated as statistically independent, we derive a general area-product rule. In 
particular,  
while the leading Bekenstein-Hawking entropy probes the sum of the horizon 
areas, the logarithmic correction probes their product. Consequently, whenever 
the classical area product is mass independent, the logarithmic entropy 
automatically inherits this universality. We demonstrate this for 
Reissner-Nordstr\"om and, under an area-based extension to rotating horizons, 
Kerr-Newman black holes. 
However, Reissner-Nordstr\"om-de Sitter provides a crucial
three-horizon counterexample, for which the area-product rule survives while
mass independence is generically lost for the physical horizons.  Thus, 
statistical 
area-product selection and classical universality are distinct properties. Our 
results uncover a structural connection between microscopic entropy counting 
and classical multi-horizon geometry that is invisible at the level of the 
leading area law, suggesting that logarithmic corrections encode distinctive 
information about the global horizon structure.

\end{abstract}

\maketitle

\section{Introduction}

Black-hole thermodynamics provides one of the most direct links between 
gravitation, quantum theory, and statistical mechanics. The Bekenstein-Hawking 
area law~\cite{Bekenstein1973,Hawking1975} suggests that the microscopic 
degrees of freedom responsible for black-hole entropy are intrinsically 
associated with horizon geometry. Beyond the leading area contribution, 
logarithmic corrections arise in a wide range of approaches, including loop 
quantum gravity, conformal and Cardy-like constructions, entanglement entropy, 
and Euclidean path-integral 
methods~\cite{
Strominger:1996sh, Ashtekar:1997yu, Kaul:1998xv, Das:2001ic, Gour:2003jj, 
KaulMajumdar2000,Perez2017,Carlip2000,Solodukhin2011,Sen2013, Mukherji:2002de, 
Govindarajan:2001ee, Strominger:1997eq, Banerjee:2010qc, Sen:2012kpz, 
Sen:2012cj, Sen:2007qy, BarberoG:2008dwr}. Although their precise form may 
depend on the underlying framework, such corrections provide a natural window 
into the microscopic structure of horizon entropy.

A seemingly different notion of universality appears in stationary black holes 
possessing more than one Killing horizon. For charged and rotating solutions, 
the areas of the outer and inner horizons can satisfy remarkable relations that 
are independent of the black-hole mass. In particular, the Kerr-Newman family 
obeys
\begin{equation}
A_+A_-=(8\pi J)^2+(4\pi Q^2)^2,
\label{eq:AH}
\end{equation}
so that the horizon-area product is determined entirely by the conserved 
angular momentum and electric charge. Such relations were established and 
explored in Refs.~\cite{AnsorgHennig2009,Cvetic:2010mn,Visser2013}, and have 
motivated the development of an inner-horizon thermodynamics in which the 
Cauchy horizon is endowed with its own thermodynamic 
structure~\cite{CastroRodriguez2012,Detournay2012,
Park:2006hu,Park:2006gt,
Birmingham:1997rj,Nian:2020qsk,Hristov:2023cuo, Cvetic:1997uw,
Chen:2012mh,Castro:2013pqa, Cvetic:2018dqf}.

The remarkable structure of horizon-area and entropy products has subsequently
been investigated for a broad variety of black-hole solutions, including charged
and rotating configurations, asymptotically flat and (anti-)de Sitter
geometries, higher-dimensional black holes, supergravity and string-inspired
solutions, and black holes in modified theories of
gravity~\cite{Chen:2012mh,Toldo:2012ec,Majeed:2015fta,Cvetic:2013eda,
Wang:2013smb,
Xu:2013zpa,Xu:2015eia,Castro:2013pqa,Cvetic:2018dqf,
Xu:2015mna,Zhang:2016nws,Cvetic:1996kv,Cvetic:1997uw,Cvetic:1997xv}.
In many cases, the resulting products exhibit a striking independence from the
mass parameter, depending instead on conserved charges, angular momenta,
cosmological constants, or gravitational couplings~\cite{Page:2015gia,
Pradhan:2013xha,Pradhan:2013hqa,MahdavianYekta:2016kqh}, although important
exceptions are also known~\cite{Debnath:2015tda}. These results have motivated
the interpretation of horizon-product relations as possible probes of the
underlying microscopic structure of black-hole
entropy~\cite{Visser:2012zi,Pradhan:2016flc,Chen:2013rb}.

Related work has also investigated logarithmic corrections to products of
inner- and outer-horizon entropies. In particular, such corrections to the
product of the corrected horizon entropies, $S_+S_-$, were found to generically
spoil the classical mass-independent structure~\cite{Pradhan:2016flc}.
The construction considered here addresses a different question, namely  rather 
than
forming a product of corrected horizon entropies, we first combine the
independent horizon sectors additively and then examine the logarithmic
contribution to the resulting total entropy, asking which combination of the
horizon areas is selected by the microscopic counting.

These two structures have rather different origins. Logarithmic entropy 
corrections concern the statistical organization of microscopic horizon degrees 
of freedom, whereas area-product relations arise from the classical geometry of 
multi-horizon solutions. This raises the central question of the present work, 
namely whether  the area-product structure, known from classical multi-horizon 
geometry, can emerge independently from microscopic entropy counting.

A particularly simple setting in which to address this question is the minimal 
statistical construction of Ref.~\cite{Calzada2026}. There, a single horizon is 
discretized into indistinguishable Planck-scale cells, and the associated Gibbs 
factor generates, through the Stirling expansion, the Bekenstein-Hawking 
contribution together with a logarithmic correction with coefficient $-1/2$. 
The original construction concerns a single connected horizon. Its extension to 
a spacetime containing several Killing horizons therefore introduces the new 
question of how   the microscopic degrees of freedom associated with 
physically distinct horizons can be statistically combined.

In this work we investigate the prescription in which the microscopic sectors 
associated with different horizons are statistically independent. This choice 
is motivated by their distinct causal and thermodynamic roles, but it is 
adopted as a working assumption rather than derived from causal separation 
itself, since correlations or entanglement between different horizon sectors 
cannot be excluded. We show that this assumption leads to a simple general 
result: while the leading Bekenstein-Hawking entropy is controlled by the 
\emph{sum} of the horizon areas, the logarithmic contribution is controlled by 
their \emph{product}. Thus, different orders in the entropy expansion probe 
different symmetric combinations of the multi-horizon geometry.

This observation establishes a direct connection with classical area-product 
universality. Whenever the product of the physical horizon areas is mass 
independent, the logarithmic entropy automatically inherits this property. We 
demonstrate this explicitly for Reissner-Nordstr\"om black holes and, under the 
extension of the area-based counting prescription to rotating horizons, for 
Kerr-Newman black holes. Importantly, we also consider Reissner-Nordstr\"om-de 
Sitter as a genuine three-horizon example. In this case the area-product rule 
continues to hold, while the product of the three physical horizon areas is 
generally mass dependent. This shows that the area-product rule is the general 
statistical result, whereas mass independence is an additional property 
inherited only when it is present in the underlying classical geometry.

More generally, non-extensive statistical mechanics provides a framework for
systems in which the standard additive composition of Boltzmann-Gibbs entropy
may cease to be appropriate, for instance in the presence of long-range
interactions, strong correlations, or other forms of nontrivial microscopic
organization \cite{ 
Tsallis:2017fhh,Tsallis:2019giw,Wilk:1999dr,Saridakis:2018unr}. Within this 
broader
framework, Tsallis and Cirto proposed a generalized non-extensive entropy
particularly motivated by gravitational systems and black-hole
thermodynamics~\cite{Tsallis:2012js}. Hence, we compare the
independent-horizon prescription with the specific Tsallis-Cirto composition
considered in Refs.~\cite{Volovik2024,Volovik2025RN,Volovik2025Kerr}. The two
approaches correspond to different assumptions about the statistical composition
and possible correlations of the microscopic degrees of freedom associated with
distinct horizons. Our aim is therefore not to claim a unique microscopic
description, but to identify the consequences of independent-horizon counting
and the unexpected connection it establishes between microscopic logarithmic
entropy and classical area-product structures.

The manuscript is organized as follows. In section \ref{sec:statistical} we 
introduce the statistical counting for multiple horizons. In section
\ref{sec:universality} we derive the general area-product rule and its 
mass-independence inheritance property. In section \ref{sec:applications} we 
apply the construction to explicit black-hole geometries and discuss its 
statistical interpretation and limitations. Finally, section \ref{Conclusions} 
contains our 
conclusions.

 \section{Statistical entropy of multiple horizons}
\label{sec:statistical}

In this section we construct the statistical framework that will be used to 
describe the entropy of spacetimes possessing more than one Killing horizon. We 
begin by briefly reviewing the single-horizon counting prescription of 
Ref.~\cite{Calzada2026}, emphasizing the origin of the logarithmic correction 
that will be crucial for our analysis. We then extend the construction to 
several horizons and discuss the statistical assumption required in order to 
combine their microscopic degrees of freedom.

\subsection{Single-horizon counting and logarithmic correction}
\label{subsec:single}

Let us first consider a single connected horizon of area $A$. Following the 
minimum-assumptions discretization  \cite{Calzada2026}, the horizon is 
divided into
\begin{equation}
N=\frac{A}{\ell_p^2} 
\label{eq:Nsingle}
\end{equation}
Planck-scale cells, where $\ell_p$ is the Planck length. The essential 
statistical assumption is that these cells are indistinguishable. Consequently, 
the corresponding phase-space volume acquires the Gibbs factor $1/N!$ and can 
be written schematically as
\begin{equation}
\Omega_N=
\frac{1}{N!}
\left(\frac{A}{L^2}\right)^N,
\label{eq:singleOmega}
\end{equation}
where $L$ denotes the characteristic length scale entering the phase-space 
construction of Ref.~\cite{Calzada2026}. The entropy is then obtained from
\begin{equation}
S=\ln\Omega_N,
\label{eq:singleSdef}
\end{equation}
together with the appropriate identification of the microscopic parameters that 
reproduces the Bekenstein-Hawking leading term.

For a macroscopic horizon, $N\gg1$, the factorial in Eq.~\eqref{eq:singleOmega} 
can be expanded through the Stirling series,
\begin{equation}
\ln N!
=
N\ln N-N+\frac{1}{2}\ln(2\pi N)
+\mathcal{O}\left(\frac{1}{N}\right).
\label{eq:Stirling}
\end{equation}
After implementing the normalization of Ref.~\cite{Calzada2026}, one obtains
\begin{equation}
S(A)
=
\frac{A}{4\ell_p^2}
-\frac{1}{2}
\ln\left(\frac{A}{\ell_p^2}\right)
+\mathcal{O}\left(\frac{\ell_p^2}{A}\right),
\label{eq:singleentropy}
\end{equation}
up to area-independent constants that are irrelevant for the considerations 
below. Hence, the standard Bekenstein-Hawking entropy appears as the dominant 
contribution, while the indistinguishability of the Planck-scale cells 
generates a logarithmic correction with coefficient $-1/2$.

The logarithmic term in Eq.~\eqref{eq:singleentropy} is of particular interest. 
Corrections of the form
\begin{equation}
S=S_{\rm BH}+c\ln S_{\rm BH}+\cdots
\label{eq:general_log}
\end{equation}
arise in several independent approaches to black-hole thermodynamics and 
quantum gravity, although the precise coefficient $c$ may depend on the 
microscopic framework, the field content and the ensemble under consideration. 
In the specific counting prescription adopted here one has $c=-1/2$, and it is 
this definite statistical structure that we will extend to the case of several 
horizons.

At this stage there is no ambiguity concerning the counting, since for a single 
connected horizon all $N$ cells belong to the same surface and are treated as 
members of one indistinguishable ensemble. The situation changes qualitatively 
when the spacetime possesses several distinct Killing horizons. In that case, 
one must additionally specify whether microscopic cells associated with 
different horizons belong to a common statistical ensemble or instead 
constitute separate statistical sectors. As we show in the following 
subsection, this apparently simple choice determines whether the logarithmic 
contribution is sensitive to the sum or to the product of the individual 
horizon 
areas.

\subsection{Independent-horizon counting}
\label{subsec:independent}

We now consider a stationary spacetime possessing $m$ distinct Killing horizons 
with areas $A_1,\ldots,A_m$. Following the single-horizon construction of the 
previous subsection, we associate with each horizon a number of Planck-scale 
cells
\begin{equation}
N_i=\frac{A_i}{\ell_p^2},
\qquad
i=1,\ldots,m.
\label{eq:Ni}
\end{equation}
The extension of the counting prescription to this case requires an additional 
assumption that is absent for a single connected horizon: one must specify how 
the microscopic degrees of freedom associated with different horizons are 
statistically combined.

One possibility is to regard all cells as belonging to a single 
indistinguishable ensemble, independently of the horizon on which they reside. 
Defining
\begin{equation}
N=\sum_{i=1}^{m}N_i,
\label{eq:Ntotal}
\end{equation}
the single-horizon prescription would then be applied to the total number of 
cells and to the total area, giving
\begin{equation}
\Omega_{\rm pooled}
=
\frac{1}{N!}
\left(
\frac{\sum_{i=1}^{m}A_i}{L^2}
\right)^N.
\label{eq:pooled}
\end{equation}
This prescription assumes that a microscopic cell associated with one horizon 
is statistically exchangeable with a cell associated with any other horizon. In 
other words, indistinguishability would be regarded as a property of the cells 
themselves, independently of the particular horizon to which they belong.

In the present work we adopt a different prescription. We assume that the 
microscopic cells are indistinguishable within each individual horizon, as in 
the single-horizon construction, while the statistical sectors associated with 
distinct horizons are independent. The total number of microstates then 
factorizes according to
\begin{equation}
\Omega_{\rm tot}
=
\prod_{i=1}^{m}\Omega_{N_i}
=
\prod_{i=1}^{m}
\frac{1}{N_i!}
\left(
\frac{A_i}{L^2}
\right)^{N_i}.
\label{eq:factorized}
\end{equation}
Consequently, the corresponding entropy is additive,
\begin{equation}
S_{\rm tot}
=
\ln\Omega_{\rm tot}
=
\sum_{i=1}^{m}\ln\Omega_{N_i}
=
\sum_{i=1}^{m}S_i.
\label{eq:Stotadditive}
\end{equation}

There are physical reasons that motivate this choice. Distinct Killing horizons 
correspond to different null surfaces and, in general, possess different 
surface gravities and temperatures. In the familiar Reissner-Nordstr\"om and 
Kerr-Newman geometries, the outer event horizon and the inner Cauchy horizon 
also play fundamentally different causal roles. In particular, only the outer 
horizon bounds the domain of outer communication, whereas the inner horizon is 
encountered only after the event horizon has been crossed. The thermodynamic 
treatment of the inner horizon as a distinct system, characterized by its own 
area, surface gravity and first-law-type relation, is also the basis of the 
inner-mechanics approach~\cite{CastroRodriguez2012,Detournay2012}. It is 
therefore natural, within the statistical framework considered here, to 
associate the different horizons with distinct microscopic sectors rather than 
to combine all their cells into a single indistinguishable ensemble. In this 
sense, statistical independence provides the minimal composition
prescription once the different horizons are regarded as distinct thermodynamic
sectors, namely  it introduces no inter-horizon correlations beyond those 
already
encoded in the classical geometry. The factorized construction should therefore
be viewed as a natural baseline against which more general correlated or
non-extensive descriptions can be compared.

Nevertheless, the statistical independence assumed in \eqref{eq:factorized} 
does not follow mathematically from causal separation alone. Distinct or 
causally separated subsystems may in principle possess correlations or quantum 
entanglement, in which case their microscopic state spaces need not factorize. 
Moreover, the counting prescription of Ref.~\cite{Calzada2026} is formulated at 
the level of phase-space counting and does not provide an underlying 
Hilbert-space structure from which possible inter-horizon correlations could be 
derived. Hence,  relation \eqref{eq:factorized} should be understood as a 
physically 
motivated working assumption of the present construction, rather than as a 
general consequence of multi-horizon geometry.

A different microscopic theory, in which the degrees of freedom associated with 
distinct horizons were correlated, could therefore lead to a different 
composition law for the entropy. The two prescriptions above should thus be 
regarded as genuinely different statistical hypotheses rather than as 
equivalent reformulations of the same counting problem. In the following 
section we examine the consequences of the independent-horizon prescription 
adopted here and show that it leads to a characteristic dependence of the 
logarithmic entropy on the geometry of the full multi-horizon system.

\section{Area-product universality}
\label{sec:universality}

Having established the statistical prescription for multiple horizons, we can
now examine its geometrical implications. The crucial consequence of
independent-horizon counting is that the leading and logarithmic contributions
to the entropy probe different combinations of the horizon areas. As we show
below, this distinction provides a direct connection between the microscopic
counting introduced above and the area-product relations that characterize
several classes of multi-horizon black holes.

\subsection{General area-product rule}
\label{subsec:generalproduct}

We now derive the main consequence of the independent-horizon counting 
prescription introduced in susbsection \ref{subsec:independent}. For a 
stationary spacetime possessing $m$ 
distinct Killing horizons with areas $A_1,\ldots,A_m$, statistical independence 
implies, according to Eq.~\eqref{eq:Stotadditive}, that the total entropy is 
the sum of the individual horizon entropies. Using the single-horizon result 
\eqref{eq:singleentropy}, and assuming $N_i=A_i/\ell_p^2\gg1$ for every 
horizon, we obtain
\begin{equation}
S_{\rm tot}
=
\sum_{i=1}^{m}
\left[
\frac{A_i}{4\ell_p^2}
-\frac{1}{2}\ln\left(\frac{A_i}{\ell_p^2}\right)
+\mathcal{O}\left(\frac{\ell_p^2}{A_i}\right)
\right].
\label{eq:Stotgeneral}
\end{equation}
Combining the logarithmic contributions, this expression becomes
\begin{equation}
S_{\rm tot}
=
\frac{1}{4\ell_p^2}\sum_{i=1}^{m}A_i
-\frac{1}{2}
\ln\left(
\frac{\prod_{i=1}^{m}A_i}{\ell_p^{2m}}
\right)
+\cdots,
\label{eq:Sproductgeneral}
\end{equation}
where the ellipsis denotes terms subleading in the large-area expansion. Thus, 
the leading and logarithmic sectors of the entropy probe two qualitatively 
different combinations of the multi-horizon geometry:
\begin{equation}
S_{\rm BH}
=
\frac{1}{4\ell_p^2}\sum_{i=1}^{m}A_i,
\qquad
S_{\log}
=
-\frac{1}{2}
\ln\left(
\frac{\prod_{i=1}^{m}A_i}{\ell_p^{2m}}
\right).
\label{eq:leadinglog}
\end{equation}
The leading Bekenstein-Hawking contribution is therefore controlled by the sum 
of the horizon areas, whereas the logarithmic correction is controlled by their 
product.

This distinction admits a useful geometrical interpretation. The horizon areas 
$A_1,\ldots,A_m$ define a set of elementary symmetric combinations, of which 
the first and the highest-order ones are
\begin{equation}
\sigma_1
=
\sum_{i=1}^{m}A_i,
\qquad
\sigma_m
=
\prod_{i=1}^{m}A_i.
\label{eq:symmetric}
\end{equation}
In terms of these quantities,  \eqref{eq:leadinglog} takes the simple form
\begin{equation}
S_{\rm BH}
=
\frac{\sigma_1}{4\ell_p^2},
\qquad
S_{\log}
=
-\frac{1}{2}
\ln\left(
\frac{\sigma_m}{\ell_p^{2m}}
\right).
\label{eq:symmetricentropy}
\end{equation}
Hence, the first two orders of the entropy expansion select different symmetric 
combinations of the same horizon geometry, namely  the extensive contribution 
probes 
$\sigma_1$, while the logarithmic contribution probes $\sigma_m$. This 
separation follows solely from the independent-horizon counting and does not 
rely on the field equations or on any particular black-hole solution.

Within the independent-horizon prescription adopted in  subsection 
\ref{subsec:independent}, this result
can be summarized in the following proposition.

\noindent\textbf{Proposition 1.}
\textit{Let a stationary spacetime possess $m$ Killing horizons of areas 
$A_1,\ldots,A_m$, each discretized into $N_i=A_i/\ell_p^2$ Planck-scale cells 
according to the counting prescription of Ref.~\cite{Calzada2026}. If the 
microscopic sectors associated with different horizons are statistically 
independent, then, for $N_i\gg1$, the logarithmic contribution to the total 
entropy is determined by the product of the horizon areas according to}
\begin{equation}
S_{\log}
=
-\frac{1}{2}
\ln\left(
\frac{\prod_{i=1}^{m}A_i}{\ell_p^{2m}}
\right).
\label{eq:boxedresult}
\end{equation}

We refer to Eq.~\eqref{eq:boxedresult} as the \emph{area-product rule} for the 
logarithmic entropy. Its statistical origin can be seen directly from the fact 
that each independent horizon contributes one term $-\frac{1}{2}\ln N_i$. 
Thus, we have
\begin{equation}
-\frac{1}{2}\sum_{i=1}^{m}\ln N_i
=
-\frac{1}{2}\ln\left(\prod_{i=1}^{m}N_i\right)
=
-\frac{1}{2}\ln\left(
\frac{\prod_{i=1}^{m}A_i}{\ell_p^{2m}}
\right).
\label{eq:logproductorigin}
\end{equation}
The area product is therefore not introduced from the known geometrical 
properties of particular black-hole solutions, but it emerges directly from the 
factorized microscopic counting. In contrast, the pooled prescription discussed 
in  subsection 
\ref{subsec:independent} would combine the horizon degrees of freedom before 
the 
logarithmic correction is generated and would therefore not produce the 
highest-order symmetric combination $\sigma_m$.

 It is worth being explicit about where the novelty of this
result lies, since the algebraic step in \eqref{eq:logproductorigin} is,
by itself, an elementary identity: $\sum_i\ln N_i=\ln(\prod_iN_i)$ holds for
any set of positive numbers, independently of any physical input. The
nontrivial content of Proposition~1 is therefore not this identity, but the
physical principle that selects it, namely the statistical independence of
 subsection 
\ref{subsec:independent}, which determines that the entropy is a \emph{sum} of 
$m$ separate
logarithms $-\frac{1}{2}\ln N_i$ in the first place, rather than a single
logarithm $-\frac{1}{2}\ln(\sum_iN_i)$ of the pooled cell number as in
Eq.~\eqref{eq:pooled}. Once independence is assumed, the appearance of the
area product is unavoidable, and the substantive physical claim is the 
assumption
itself, not the elementary algebra that follows from it. 

It should be stressed that Proposition~1 does not explain the geometric origin 
of classical area-product identities. Rather, it establishes that the 
statistical construction independently selects the same type of horizon 
invariant. This observation becomes particularly significant when $\sigma_m$ 
itself possesses universal properties. In particular, if the classical area 
product is independent of the black-hole mass,  relation \eqref{eq:boxedresult} 
implies that the logarithmic entropy inherits the same property. 

 It is natural to ask about the fate of the intermediate
elementary symmetric combinations $\sigma_2,\ldots,\sigma_{m-1}$ for $m>2$,
which do not appear in \eqref{eq:symmetricentropy}. Within the present
counting, they simply do not arise at this order. Namely, the expansion in
expression \eqref{eq:Stotgeneral} is a sum of single-horizon contributions, 
each
depending only on its own $A_i$, so only the fully symmetric extremes
$\sigma_1$ (from the linear terms) and $\sigma_m$ (from the sum of logarithms)
appear at the orders considered. Intermediate symmetric combinations could in
principle enter through cross terms if the independent-horizon factorization
were relaxed, but they are absent order by order in the present, strictly
factorized construction. 

\subsection{Inheritance of mass independence}
\label{subsec:inheritance}

The area-product rule derived above acquires a stronger significance for 
black-hole families whose horizon areas satisfy a universal product relation. 
In several stationary multi-horizon solutions, the product of the horizon areas 
can be expressed entirely in terms of conserved charges, angular momenta, and 
other parameters characterizing the solution, while being independent of the 
mass. Let us consider, in general, a family for which
\begin{equation}
\prod_{i=1}^{m}A_i
=
\mathcal{I}(Q_a,J_a,\ldots),
\label{eq:generalinvariant}
\end{equation}
with
\begin{equation}
\frac{\partial \mathcal{I}}{\partial M}=0,
\label{eq:massindependentI}
\end{equation}
where the derivative is understood at fixed conserved charges, angular momenta, 
and any other parameters entering $\mathcal{I}$.
Substituting \eqref{eq:generalinvariant} into the area-product rule 
\eqref{eq:boxedresult}, we obtain
\begin{equation}
S_{\log}
=
-\frac{1}{2}
\ln\left[
\frac{\mathcal{I}(Q_a,J_a,\ldots)}
{\ell_p^{2m}}
\right],
\label{eq:Sloginvariant}
\end{equation}
and it follows immediately that
\begin{equation}
\frac{\partial S_{\log}}{\partial M}
=
-\frac{1}{2}
\frac{1}{\mathcal{I}}
\frac{\partial \mathcal{I}}{\partial M}
=
0.
\label{eq:Slogmassindependent}
\end{equation}
Hence, whenever the classical horizon geometry possesses a mass-independent 
area product, the logarithmic contribution obtained from independent-horizon 
counting is itself exactly mass independent.

This result can be stated as a direct consequence of Proposition~1.

\noindent\textbf{Corollary 1 (inheritance of mass independence).}
\textit{Consider a stationary spacetime with $m$ Killing horizons whose 
microscopic sectors satisfy the independent-horizon counting prescription of 
 subsection 
\ref{subsec:independent}. If the corresponding classical horizon areas obey}
\begin{equation}
\prod_{i=1}^{m}A_i
=
\mathcal{I}(Q_a,J_a,\ldots),
\qquad
\frac{\partial\mathcal{I}}{\partial M}=0,
\label{eq:inheritancecondition}
\end{equation}
\textit{then the logarithmic contribution to the total entropy inherits the 
same mass independence,}
\begin{equation}
S_{\log}
=
-\frac{1}{2}
\ln\left[
\frac{\mathcal{I}(Q_a,J_a,\ldots)}
{\ell_p^{2m}}
\right],
\qquad
\frac{\partial S_{\log}}{\partial M}=0.
\label{eq:inheritance}
\end{equation}

We refer to this result as the \emph{inheritance property} of the logarithmic 
entropy. Its significance lies in the fact that the two ingredients entering 
the argument have independent origins. The relation \eqref{eq:generalinvariant} 
is a property of the classical horizon geometry, determined by the 
corresponding gravitational solution. In contrast, the dependence of $S_{\log}$ 
on the area product follows from the microscopic statistical counting and, 
specifically, from the factorization of the horizon sectors. Thus, mass 
independence is not imposed by the statistical construction itself, but  
rather, 
once the area product is selected by the counting, the logarithmic entropy 
inherits the universality properties already possessed by that product.

As seen in subsection \ref{subsec:generalproduct}, the leading and logarithmic 
contributions probe 
different symmetric combinations of the horizon areas. Consequently, even when 
the highest-order combination $\sigma_m$ is independent of $M$, the first 
combination $\sigma_1$ need not be. The entropy expansion can therefore 
separate different pieces of information contained in the same multi-horizon 
geometry. The leading term may retain an explicit dependence on the mass, 
whereas the logarithmic term may depend only on conserved quantities.

The inheritance property should not be interpreted as a microscopic derivation
of the classical area-product identity. Rather, it expresses a structural
compatibility between two independently defined ingredients, i.e. the 
statistical
counting selects the area product, while the classical geometry determines
whether that product is mass independent.

Conversely, Proposition~1 remains valid within the present counting 
prescription even for black holes whose physical area product depends 
explicitly on the mass. In such cases the logarithmic entropy simply inherits 
that mass dependence, and no relation of the form 
\eqref{eq:Slogmassindependent} follows. The area-product rule is therefore the 
general statistical statement, while mass independence is an additional 
property 
that arises only when it is already present in the classical multi-horizon 
geometry.

\section{Black-hole applications and physical implications}
\label{sec:applications}

We now apply the general results of the previous section to explicit 
multi-horizon 
black-hole geometries. We begin with the Reissner-Nordstr\"om solution, where 
the distinction between the area sum and the area product can be exhibited in 
its simplest form, and then extend the analysis to rotating black holes and to 
configurations involving more than two horizons.

\subsection{Reissner-Nordstr\"om black holes}
\label{subsec:RN}

The Reissner-Nordstr\"om solution provides the simplest setting in which the 
inheritance property can be illustrated explicitly. The metric function is
\begin{equation}
f(r)
=
1-\frac{2M}{r}+\frac{Q^2}{r^2},
\label{eq:RNmetricfunction}
\end{equation}
and, for $M^2>Q^2$, it possesses an outer event horizon and an inner Cauchy 
horizon located at
\begin{equation}
r_{\pm}
=
M\pm\sqrt{M^2-Q^2}.
\label{eq:RNhorizons}
\end{equation}
The two horizon radii satisfy
\begin{equation}
r_+ + r_-
=
2M,
\qquad
r_+r_-
=
Q^2.
\label{eq:RNradiusrelations}
\end{equation}
Since
\begin{equation}
A_{\pm}=4\pi r_{\pm}^2,
\label{eq:RNareasdef}
\end{equation}
the first and highest-order elementary symmetric combinations of the two 
horizon areas are
\begin{equation}
A_+ + A_-
=
8\pi\left(2M^2-Q^2\right),
\label{eq:RNareasum}
\end{equation}
and
\begin{equation}
A_+A_-
=
16\pi^2Q^4.
\label{eq:RNareaproduct}
\end{equation}
The contrast is immediate. The area sum depends explicitly on the black-hole 
mass, whereas the area product depends only on the conserved electric charge. 
Equation~\eqref{eq:RNareaproduct} is the nonrotating limit of the universal 
inner-outer horizon area-product relation and provides a particularly 
transparent example of the mass-independent structure discussed in 
section \ref{sec:universality} 
\cite{AnsorgHennig2009,Cvetic:2010mn,Visser2013}.

Applying  \eqref{eq:Sproductgeneral} for $m=2$, the entropy associated with 
the two statistically independent horizon sectors takes the form
\begin{equation}
S_{\rm tot}^{\rm RN}
=
\frac{A_++A_-}{4\ell_p^2}
-\frac{1}{2}
\ln\left(
\frac{A_+A_-}{\ell_p^4}
\right)
+\cdots.
\label{eq:RNStot}
\end{equation}
Using Eqs.~\eqref{eq:RNareasum} and \eqref{eq:RNareaproduct}, this becomes
\begin{equation}
S_{\rm tot}^{\rm RN}
=
\frac{2\pi}{\ell_p^2}
\left(2M^2-Q^2\right)
-\frac{1}{2}
\ln\left(
\frac{16\pi^2Q^4}{\ell_p^4}
\right)
+\cdots.
\label{eq:RNStotexplicit}
\end{equation}
Thus, the separation between the leading and logarithmic sectors is explicit, 
i.e 
\begin{equation}
S_{\rm BH}^{\rm RN}
=
\frac{2\pi}{\ell_p^2}
\left(2M^2-Q^2\right),
\qquad
S_{\log}^{\rm RN}
=
-\frac{1}{2}
\ln\left(
\frac{16\pi^2Q^4}{\ell_p^4}
\right),
\label{eq:RNentropysectors}
\end{equation}
and consequently 
\begin{equation}
\frac{\partial S_{\rm BH}^{\rm RN}}{\partial M}
=
\frac{8\pi M}{\ell_p^2},
\qquad
\frac{\partial S_{\log}^{\rm RN}}{\partial M}
=
0.
\label{eq:RNmassderivatives}
\end{equation}

Hence, Reissner-Nordstr\"om provides the simplest explicit realization of 
Corollary~1. The leading entropy retains the mass dependence contained in 
$A_++A_-$, whereas the logarithmic contribution inherits the mass independence 
of the classical product $A_+A_-=16\pi^2Q^4$. This makes transparent how the 
statistical selection of the area product and the independently known classical 
horizon relation combine to yield a charge-dependent but mass-independent 
logarithmic sector.

\subsection{Kerr-Newman black holes}
\label{subsec:KN}

We next consider the Kerr-Newman family, which provides the natural rotating 
extension of the Reissner-Nordstr\"om case. The outer and inner horizons are 
located at
\begin{equation}
r_{\pm}
=
M\pm\sqrt{M^2-a^2-Q^2},
\label{eq:KNhorizons}
\end{equation}
where
\begin{equation}
a=\frac{J}{M}.
\label{eq:KNa}
\end{equation}
The corresponding horizon areas are
\begin{equation}
A_{\pm}
=4\pi\left(r_{\pm}^2+a^2\right).
\label{eq:KNareas}
\end{equation}
Although the individual areas depend explicitly on the mass, their product 
satisfies the universal relation
\begin{equation}
A_+A_-
=
(8\pi J)^2+(4\pi Q^2)^2
=
16\pi^2\left(4J^2+Q^4\right),
\label{eq:KNareaproduct}
\end{equation}
which is independent of $M$~\cite{AnsorgHennig2009,Cvetic:2010mn,Visser2013}.

Assuming that the area-based single-horizon counting underlying Sec. 
\ref{sec:statistical}
continues to apply to rotating horizons, and that the inner and outer horizon
sectors remain statistically independent, Proposition~1 gives
\begin{equation}
S_{\log}^{\rm KN}
=
-\frac{1}{2}
\ln\left[
\frac{16\pi^2\left(4J^2+Q^4\right)}
{\ell_p^4}
\right].
\label{eq:KNlog}
\end{equation}
Hence,
\begin{equation}
\frac{\partial S_{\log}^{\rm KN}}{\partial M}=0,
\label{eq:KNmassindependent}
\end{equation}
and the logarithmic contribution is controlled entirely by the conserved 
angular momentum and electric charge. Thus, Kerr-Newman provides a second 
explicit realization of the inheritance property, with the relevant classical 
invariant determined by the combination $4J^2+Q^4$.

Note that the limiting cases are recovered smoothly. For $J\to0$ one obtains 
the 
Reissner-Nordstr\"om result, namely 
\begin{equation}
S_{\log}^{\rm KN}
\longrightarrow
-\frac{1}{2}
\ln\left(
\frac{16\pi^2Q^4}{\ell_p^4}
\right),
\label{eq:KNtoRN}
\end{equation}
whereas for $Q\to0$ one obtains the Kerr limit 
\begin{equation}
S_{\log}^{\rm Kerr}
=
-\frac{1}{2}
\ln\left(
\frac{64\pi^2J^2}{\ell_p^4}
\right).
\label{eq:Kerrlog}
\end{equation}
These limits show explicitly that the mass independence of the logarithmic 
sector is not tied to electric charge alone, but follows from the conserved 
quantities entering the corresponding classical area product.

There is, however, an additional assumption involved in this application. The 
microscopic phase-space construction of Ref.~\cite{Calzada2026} was formulated 
for static, spherically symmetric horizons, and a corresponding 
first-principles derivation for rotating horizons has not yet been established. 
In particular, frame dragging may modify the local phase-space structure 
associated with the horizon cells. Therefore, Eq.~\eqref{eq:KNlog} should be 
understood as an extension of the area-based counting prescription under the 
assumption that the same horizon-area dependence remains valid in the rotating 
case. This assumption is geometrically natural, since the derivation of
Proposition~1 depends only on the horizon areas once the single-horizon
logarithmic term is given, but it should not be confused with an independent
microscopic derivation for Kerr-Newman. 
Accordingly, the Kerr-Newman result
should be regarded as a conditional extension of the general area-product rule,
rather than as a result established on the same microscopic footing as the
static Reissner-Nordstr\"om case.

Independent calculations of logarithmic corrections to Kerr thermodynamics
through Euclidean and near-horizon methods~\cite{KapecKerrLog2023} may provide
useful benchmarks for a future microscopic extension of the present counting
scheme to rotating horizons.

\subsection{Beyond two horizons}
\label{subsec:beyondtwo}

The previous examples involve two physical horizons. However, Proposition~1 was 
formulated for an arbitrary number $m$ of Killing horizons, and it is therefore 
important to examine whether the same reasoning remains meaningful in 
geometries with a richer horizon structure. A particularly instructive example 
is provided by the Reissner-Nordstr\"om-de Sitter (RNdS) spacetime, which can 
possess 
three distinct positive horizons: an inner Cauchy horizon, a black-hole event 
horizon, and a cosmological horizon.

The metric function is
\begin{equation}
f(r)
=
1-\frac{2M}{r}
+\frac{Q^2}{r^2}
-\frac{\Lambda r^2}{3},
\label{eq:RNdSmetric}
\end{equation}
with $\Lambda>0$. The horizon equation $f(r)=0$ can be written as
\begin{equation}
r^4
-\frac{3}{\Lambda}r^2
+\frac{6M}{\Lambda}r
-\frac{3Q^2}{\Lambda}
=
0.
\label{eq:RNdSquartic}
\end{equation}
In the parameter region where three positive horizons exist, we denote them by
\begin{equation}
r_-<r_+<r_c,
\end{equation}
corresponding respectively to the Cauchy, event, and cosmological horizons. The 
fourth root, which we denote by $r_v$, is negative and does not represent a 
physical Killing horizon.

If the three physical horizons are treated as statistically independent 
sectors, Proposition~1 gives
\begin{equation}
S_{\log}^{\rm RNdS}
=
-\frac{1}{2}
\ln\left(
\frac{A_-A_+A_c}{\ell_p^6}
\right),
\label{eq:RNdSlog}
\end{equation}
where
\begin{equation}
A_i=4\pi r_i^2.
\end{equation}
This is a genuine $m=3$ realization of the general area-product rule, i.e. the 
logarithmic entropy is controlled by the product of all three physical horizon 
areas rather than by their sum.

Before
proceeding we should mention   a caveat specific to this case.   Treating 
the cosmological horizon on exactly the same statistical
footing as the black-hole event and Cauchy horizons, is a further extension of
susbsection \ref{subsec:independent}, beyond the extension to rotating horizons 
already noted in
subsection \ref{subsec:KN}. The cosmological horizon of de~Sitter-type 
spacetimes raises
well-known conceptual issues not shared by black-hole horizons, such as observer
dependence, the absence of a global timelike Killing vector covering the
entire spacetime, and the restriction to a single static patch, which have
motivated a long-standing and still unsettled debate about the precise
statistical meaning of de~Sitter entropy. Relation \eqref{eq:RNdSlog} should
accordingly be read as a further, and comparatively more speculative,
application of the independent-horizon prescription, on top of the
rotating-horizon extension of subsection \ref{subsec:KN}, rather than as being 
established on the same footing as the
Reissner-Nordstr\"om case of   subsection \ref{subsec:RN}. 

The Reissner-Nordstr\"om-de Sitter case, however, also illustrates an 
important distinction between the area-product rule itself and the inheritance 
of mass independence. From Vieta's relations applied to 
Eq.~\eqref{eq:RNdSquartic}, the product of all four roots is
\begin{equation}
r_-r_+r_cr_v
=
-\frac{3Q^2}{\Lambda}.
\label{eq:RNdSrootproduct}
\end{equation}
Consequently, the product of the areas associated with all four algebraic roots 
is
\begin{equation}
A_-A_+A_cA_v
=
(4\pi)^4
\left(
r_-r_+r_cr_v
\right)^2
=
\frac{2304\pi^4Q^4}{\Lambda^2},
\label{eq:RNdSallproduct}
\end{equation}
which is exactly independent of the mass. Such mass-independent products 
involving all roots, including virtual horizons, are familiar from the study of 
multi-horizon area and entropy relations 
\cite{Cvetic:2010mn,Cvetic:2013eda,Wang:2013smb,Xu:2013zpa,Wang:2013nvz,
Xu:2014qaa,Xu:2014qza,Xu:2015eia,Xu:2015mna,Du:2014kpa,Zhang:2014kna,
Liu:2016ckh,
Zhang:2016nws,Cvetic:2018dqf,Pradhan:2015wnl,Pradhan:2020ofm}.

For the present statistical construction, however, the relevant objects are the 
physical horizons. Eliminating the virtual root from 
Eq.~\eqref{eq:RNdSrootproduct} gives
\begin{equation}
r_-r_+r_c
=
-\frac{3Q^2}{\Lambda r_v},
\label{eq:RNdSphysicalrootproduct}
\end{equation}
and therefore
\begin{equation}
A_-A_+A_c
=
(4\pi)^3
\left(
\frac{3Q^2}{\Lambda r_v}
\right)^2
=
\frac{576\pi^3Q^4}{\Lambda^2 r_v^2}.
\label{eq:RNdSphysicalareaproduct}
\end{equation}
Since the virtual root $r_v$ depends in general on the mass parameter through 
Eq.~\eqref{eq:RNdSquartic}, the product of the three physical horizon areas is 
not generically mass independent. Accordingly,
\begin{equation}
\frac{\partial}{\partial M}
\left(
A_-A_+A_c
\right)
\neq0,
\end{equation}
and hence, in general,
\begin{equation}
\frac{\partial S_{\log}^{\rm RNdS}}{\partial M}
\neq0.
\label{eq:RNdSmassdependent}
\end{equation}

Thus, independent-horizon counting continues to imply the area-product rule,
\begin{equation}
S_{\log}
\propto
-\ln\left(\prod_i A_i\right),
\label{eq:RNdSgeneralrule}
\end{equation}
while mass independence is not guaranteed. The Reissner-Nordstr\"om-de Sitter
case therefore provides an explicit example in which Proposition~1 remains
valid for the physical horizons, whereas Corollary~1 does not apply.

The virtual root introduces an additional conceptual distinction. If one
formally included all four roots in the statistical expression,
Eq.~\eqref{eq:RNdSallproduct} would yield
\begin{equation}
S_{\log}^{\rm all}
=
-\frac{1}{2}
\ln\left(
\frac{2304\pi^4Q^4}
{\Lambda^2\ell_p^8}
\right),
\label{eq:RNdSalllog}
\end{equation}
which is again mass independent. However, such an extension has no direct
justification within the microscopic construction adopted here, since the
negative root does not correspond to a physical horizon that can naturally be
associated with an independent ensemble of Planck-scale cells. Hence, although
the algebraic product over all roots possesses a universal mass-independent
form, the corresponding statistical interpretation is meaningful only for
physical horizons. This distinction shows that geometric universality alone is
not sufficient, and  the horizons entering the product must also admit a 
consistent
microscopic interpretation.
 
The Reissner-Nordstr\"om-de Sitter example therefore provides both a genuine
extension of the area-product rule beyond two horizons and a useful counterpoint
to the asymptotically flat Reissner-Nordstr\"om and Kerr-Newman cases. In
particular, it delineates the domain of the inheritance property rather than
merely providing another realization of it. For transparency, the different 
cases considered
above, and the distinction between the general area-product rule and the
additional inheritance of mass independence, are summarized in
Table~\ref{tab:summary}.

\begin{table*}[t]
\centering
\caption{Summary of the area-product rule and the inheritance property for the
black-hole geometries considered in this work. The logarithmic entropy is
controlled by the product of the physical horizon areas, whereas mass
independence is inherited only when this physical area product is itself
independent of the mass. For Reissner-Nordstr\"om-de Sitter (RNdS), the product 
over all four algebraic roots is
also shown for comparison. The fourth root is negative and does not correspond
to a physical Killing horizon, hence, although the corresponding algebraic area
product is mass independent, it is not included in the statistical counting
adopted here.}
\label{tab:summary}
\begin{ruledtabular}
\begin{tabular}{lcccc}
Geometry
& Horizons
& Area product
& Mass independent
& Inheritance
\\
\hline
Reissner-Nordstr\"om
& $2$
& $A_+A_-=16\pi^2Q^4$
& Yes
& Yes
\\
Kerr-Newman
& $2$
& $A_+A_-=16\pi^2(4J^2+Q^4)$
& Yes
& Yes
\\
Reissner-Nordstr\"om-de Sitter
& $3$
& $A_-A_+A_c$
& No, generically
& No
\\
RNdS (all algebraic roots)
& $4$
& $A_-A_+A_cA_v=2304\pi^4Q^4/\Lambda^2$
& Yes
& Not applicable
\end{tabular}
\end{ruledtabular}
\end{table*}

\subsection{Statistical interpretation and limitations}
\label{subsec:interpretation}

The results derived above rely on a specific statistical interpretation of the
microscopic degrees of freedom associated with different horizons, namely  
that the
factorization $\Omega_{\rm tot}=\prod_i\Omega_i$  and hence the additive
composition $S_{\rm tot}=\sum_iS_i$, correspond to treating the different
horizon sectors as statistically independent. This is a physically motivated
prescription, but it is not unique. Indeed, a variety of generalized and
non-extensive entropy frameworks have been proposed, including Tsallis,
Kaniadakis, and Luciano-Saridakis entropies, which modify the standard
Boltzmann-Gibbs statistical structure and may lead to non-additive entropy
composition 
laws~\cite{Tsallis:2017fhh,Tsallis:2019giw,Wilk:1999dr,Saridakis:2018unr,
Ghaffari:2018wks, Aditya:2019bbk, 
Nojiri:2021czz,Kaniadakis:2002zz,Kaniadakis:2005zk,Drepanou:2021jiv,
Lymperis:2021qty,Luciano:2026ufu,Anand:2025rjg,Anand:2025cer,
Leizerovich:2026pfy, Luciano:2026eiy}. Such
frameworks illustrate more generally that the thermodynamic description of
gravitational systems can depend sensitively on the underlying statistical
prescription.
 
In particular, an instructive alternative in the specific context of 
multi-horizon black
holes is provided by the non-extensive Tsallis-Cirto composition considered in
Refs.~\cite{Volovik2024,Volovik2025RN,Volovik2025Kerr}. For 
Reissner-Nordstr\"om, the outer horizon is assigned
\begin{equation}
S_+
=
\frac{\pi r_+^2}{\ell_p^2},
\end{equation}
while the inner horizon carries
\begin{equation}
S_-
=
-\frac{\pi r_-^2}{\ell_p^2},
\end{equation}
and the two are combined according to
\begin{equation}
\sqrt{S_{\rm RN}}
=
\sqrt{S_+}
+
\sqrt{|S_-|}.
\label{eq:TsallisCirto}
\end{equation}
Using $r_++r_-=2M$, this yields
\begin{equation}
S_{\rm RN}
=
\frac{4\pi M^2}{\ell_p^2},
\label{eq:TsallisRN}
\end{equation}
which depends only on the mass and carries no residual charge dependence.

This structure differs fundamentally from the independent-horizon prescription 
adopted here, for which the horizon entropies are combined extensively and the 
logarithmic sector of Reissner-Nordstr\"om depends only on the charge. The two 
approaches should therefore not be viewed as competing calculations within the 
same statistical framework, but as different assumptions about how the 
microscopic sectors associated with distinct horizons are composed. The 
distinction may ultimately reflect the role of inter-horizon correlations: 
factorization is natural for independent microscopic sectors, whereas 
sufficiently strong correlations or entanglement could invalidate an extensive 
composition law. Neither framework presently provides a microscopic derivation 
of such correlations, and hence the appropriate composition rule remains an 
additional physical input.

It is also important to clarify the meaning of $S_{\rm tot}$ in the present 
construction. We do not propose it as a replacement for the standard 
Bekenstein-Hawking entropy of the outer event horizon entering the usual first 
law and describing thermodynamics accessible to an asymptotic observer. Rather, 
$S_{\rm tot}$ is the combinatorial entropy obtained under the assumption that 
each physical Killing horizon carries its own microscopic degrees of freedom. 
In this respect, the construction is closer in spirit to the inner-mechanics 
program~\cite{CastroRodriguez2012,Detournay2012}, in which the inner horizon is 
endowed with an independent thermodynamic structure.

A further limitation follows from the large-area approximation underlying the 
counting. The Stirling expansion requires
\begin{equation}
N_i
=
\frac{A_i}{\ell_p^2}
\gg1
\end{equation}
for every horizon included in the ensemble. This condition breaks down toward 
the Schwarzschild limit of Reissner-Nordstr\"om, where $A_-\to0$. The extremal 
regime likewise requires separate treatment, namely  as the inner and outer 
horizons 
merge, they can no longer be regarded as distinct statistical sectors. Thus, 
although the formal area-product expressions remain finite at extremality, 
their interpretation in terms of independent horizon ensembles ceases to apply. 
This is consistent with the special treatment known to be required for 
logarithmic corrections in near-extremal 
geometries~\cite{IliesiuMurthyTuriaci2022}.

The significance of the mass-independent logarithmic contribution is primarily 
conceptual rather than observational. For a macroscopic black hole we have
\begin{equation}
\ln\left(\frac{A}{\ell_p^2}\right)
\ll
\frac{A}{\ell_p^2},
\end{equation}
so that the logarithmic term is overwhelmingly suppressed relative to the 
leading Bekenstein-Hawking contribution and is unlikely to be accessible 
through foreseeable black-hole observations. Its interest instead lies in the 
structural correspondence identified above, namely  within the present counting 
prescription the same area product that appears as a classical horizon 
invariant also controls the logarithmic entropy.

This correspondence should be regarded as a compatibility between the 
statistical and geometric descriptions, not as a microscopic explanation of 
classical area-product universality. Whether it reflects a deeper property of 
multi-horizon microphysics can only be determined within a framework in which 
correlations between the degrees of freedom associated with different horizons 
are derived rather than assumed.

\section{Conclusions}
\label{Conclusions}

In this work we investigated the statistical structure of black-hole entropy in 
spacetimes possessing multiple Killing horizons. Our motivation was the 
coexistence of two apparently distinct structures in black-hole physics. On the 
one hand, microscopic and semiclassical approaches generically predict 
logarithmic corrections to the Bekenstein-Hawking entropy. On the other hand, 
the classical geometry of several stationary black holes exhibits remarkable 
area-product relations, which in important cases are independent of the 
black-hole mass. We have shown that these two structures can be connected in a 
simple and nontrivial way when the microscopic degrees of freedom associated 
with distinct horizons are treated as statistically independent sectors.

Starting from the minimum-assumptions Planck-cell discretization of 
Ref.~\cite{Calzada2026}, we extended the single-horizon counting prescription 
to stationary spacetimes containing an arbitrary number of physical Killing 
horizons. The essential additional assumption is the statistical factorization 
of the corresponding microscopic sectors. Under this assumption, the total 
entropy is additive, but its leading and logarithmic contributions probe 
qualitatively different combinations of the multi-horizon geometry. In 
particular,  the 
Bekenstein-Hawking term is controlled by the sum of the horizon areas, whereas 
the logarithmic correction is controlled by their product. We formulated this 
result as the \emph{area-product rule}. Importantly, it follows directly from 
statistical factorization and does not rely on the gravitational field 
equations or on the properties of any particular black-hole solution.

A particularly interesting consequence arises when the classical product of the 
physical horizon areas is itself universal and independent of the black-hole 
mass. In this case, the logarithmic entropy automatically inherits the same 
mass independence. This \emph{inheritance property} establishes the central 
connection uncovered in the present work, namely that  the geometrical quantity 
selected by 
the microscopic statistical counting is precisely the combination that, for 
important classes of multi-horizon black holes, possesses a distinguished 
classical universality. The statistical construction does not explain why the 
classical area-product relations hold, but rather  the nontrivial observation 
is 
that two logically independent constructions single out the same combination of 
horizon data.

We demonstrated this mechanism explicitly for Reissner-Nordstr\"om black holes. 
While the leading entropy retains an explicit dependence on the mass, the 
logarithmic contribution depends only on the conserved electric charge and is 
exactly mass independent. The same structure extends to Kerr-Newman black 
holes, where the logarithmic sector is determined by the conserved angular 
momentum and electric charge while remaining independent of the mass. The 
rotating case, however, involves an additional assumption, since the underlying 
microscopic phase-space construction has not yet been derived from first 
principles for rotating horizons. Our Kerr-Newman result should therefore be 
understood as an area-based extension of the statistical prescription rather 
than as an independent microscopic derivation.

The Reissner-Nordstr\"om-de Sitter case provides an important complementary 
result and clarifies the generality of our conclusions. It constitutes a 
genuine three-physical-horizon realization of the area-product rule, but the 
product of the physical horizon areas is generically mass dependent. Thus, the 
statistical area-product rule survives, whereas the inheritance of mass 
independence does not. Interestingly, mass independence is recovered 
algebraically if the additional negative root is included, but such a root does 
not correspond to a physical horizon that can naturally be assigned an 
independent Planck-cell ensemble. This example therefore demonstrates that the 
area-product rule and mass independence are conceptually distinct. The former 
is a consequence of the statistical prescription, while the latter requires an 
additional property of the physical horizon geometry.

We have also emphasized that statistical independence is a physical assumption 
rather than an unavoidable consequence of the existence of distinct horizons. 
Alternative composition prescriptions can lead to qualitatively different 
results. In particular, comparison with the non-extensive Tsallis-Cirto 
construction shows that the two approaches represent different statistical 
descriptions of the microscopic horizon sectors rather than competing 
calculations of the same quantity.   Moreover, the present construction relies 
on a large-area expansion for every horizon and consequently requires separate 
treatment when an inner horizon becomes microscopic or when distinct horizons 
merge in the extremal limit.

The significance of our results is primarily conceptual rather than 
observational, since logarithmic corrections are overwhelmingly suppressed 
relative to the leading entropy for macroscopic black holes. Nevertheless, they 
provide a sensitive probe of the statistical organization of microscopic 
horizon degrees of freedom. In this respect, the fact that independent-horizon 
counting naturally selects the same area-product structure that appears 
independently as a distinguished invariant of classical multi-horizon geometry 
is particularly suggestive. It indicates a structural compatibility between the 
microscopic statistical description and the classical geometry that is not 
apparent at the level of the leading area law alone.

Several directions deserve further investigation. A first priority is to
develop a microscopic framework capable of determining the correlations and
entanglement between distinct horizon sectors, thereby allowing the statistical
composition law itself to be derived rather than assumed. In particular,
deriving, or falsifying, the factorization of the horizon sectors would provide
a decisive test of the statistical premise underlying the area-product rule
obtained here. Recent studies of the quantum structure of Cauchy horizons and
of holographic complexity growth across inner horizons may offer promising
avenues in this direction, since a first-principles characterization of
inter-horizon correlations could help distinguish between the 
independent-horizon
prescription adopted here and possible non-extensive alternatives.

A further important direction is to formulate the microscopic phase-space
counting directly for rotating horizons, thereby testing the Kerr-Newman
extension from first principles. It would also be interesting to extend the
analysis to higher-dimensional, regular, and modified-gravity black holes,
where the structure and universality of horizon-area products can be
considerably richer. Finally, extremal and near-extremal configurations require
a dedicated treatment beyond the independent large-area ensembles considered
here. These developments may ultimately reveal whether the correspondence
uncovered in this work between logarithmic entropy and classical area-product
invariants is specific to the present counting prescription or reflects a more
general organizing principle of multi-horizon black-hole microphysics.

\begin{acknowledgments}

Y. G\'omez-Leyton acknowledges support from the ANID ``Subvenci\'on en la
Academia'' program, 2025 call, Grant No.~85250186.  E.N. Saridakis 
acknowledges the contribution of the LISA   CosWG, and     COST   
Actions  
 CA21106 ``COSMIC WISPers
in the Dark Universe: Theory, astrophysics and experiments'',  CA21136 
``Addressing observational tensions in cosmology with 
  systematics and fundamental physics (CosmoVerse)'',    CA23130 
``Bridging high and low energies in
search of quantum gravity (BridgeQG)'', and CA24101 ``Testing Fundamental 
Physics with Seismology''.
\end{acknowledgments}

\bibliography{references.bib}

\end{document}